# Cough Detection from Acoustic signals for patient monitoring system

Vinay Kulkarni , Radhakrishnan Vadakkethil (Ignitarium Technology Solutions Pvt. Ltd)

*Abstract*— Cough is one of the most common symptoms in all respiratory diseases. In cases like Chronic Obstructive Pulmonary Disease, Asthma, acute and chronic Bronchitis and the recent pandemic Covid-19, the early identification of cough is important to provide healthcare professionals with useful clinical information such as frequency, severity, and nature of cough to enable better diagnosis. This paper presents and demonstrates best feature selection using MFCC which can help to determine cough events, eventually helping a neural network to learn and improve accuracy of cough detection. The paper proposes to achieve performance of 97.77% Sensitivity (SE), 98.75% Specificity (SP) and 98.17% F1-score with a very light binary classification network of size close to 16K parameters, enabling fitment into smart IoT devices.

*Keywords—Cough detection, MFCC, Convolutional Neural Network (CNN), IoT Edge devices, patient monitoring systems.*

## I. Introduction

In the era of IoT devices, smart patient monitoring systems are evolving with real time monitoring, which are deployable at home and hospitals at reduced cost. One such application is cough sound detection and classification as a part of smart home or smart patient monitoring system. The Cough is one of the most common symptoms in all respiratory diseases, and their characteristics helps in early identification and management of diseases. During Chronic Respiratory Disease (CRD) such as asthma, and Chronic Obstructive Pulmonary Disease (COPD), the occurrences of coughs and onset of their spells along with number of cough events help healthcare professionals to understand the progression of the disease, thereby helping in treatment of the same [1]. Describing the frequency, severity and nature of cough is challenging for the aged, Parkinson's or Alzheimer's patients, and Covid-19 patients. Many a time, tracking the frequency and severity of cough during sleep is also challenging, which leads us to bank on smart monitoring systems.

To leverage Machine learning capabilities, the neural network is often selected to detect the cough sounds. Deep Neural Network (DNN) for classification with Mel-Frequency Cepstral Coefficient (MFCC) as feature inputs was used by papers, with performance of 90.1% SE and 85% SP[3], 86.8% SE and 92.7% SP[4], 89.7% SE and 89.85% SP[5], 92.8% SE and 97.5% SP[6]. The last one used MFCC, zero crossing rate and Shannon entropy to achieve best performance. The addition of more features and deeper networks add load on IoT edge devices, which must be considered while developing smart patient monitoring systems.

The paper introduces cough sound characteristics and their feature uniqueness in Part II, along with MFCC feature. Part-III covers the development of cough detections. Part-IV touches upon database, performance, and device resources/footprints. Part-V ends with conclusions.

## II. Cough sound characteristics and MFCC

Cough is described as "a three-phase expulsive motor act characterized by an aspiratory effort (aspiratory phase) followed by a forced expiratory effort initially against a closed glottis (compressive phase), and then by active glottal opening and rapid expiratory flow (expulsive phase)" [7]. These will be referenced as Phase1,Phase2 and Phase3 respectively in this paper. There are some cases in which Phase 3 is not visible in the cough signal [8]. The Fig.1 shows typical cough sound characteristics and their respective log spectrogram. The upper part of Fig.1 shows demarcation of Phase-1, Phase-2 and Phase-3 in time domain waveform.

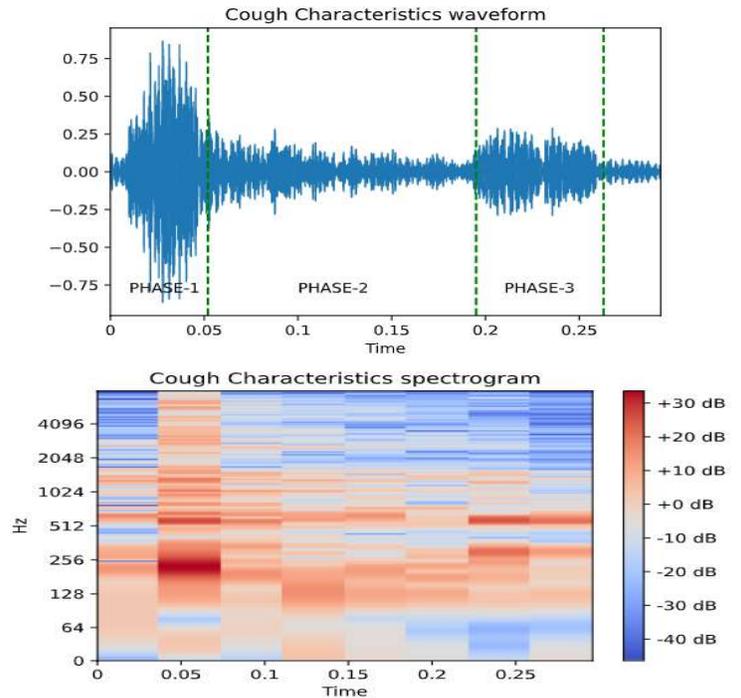

Fig. 1: Cough characteristics

The cough sound typically varies from 0.25 sec to 1 sec duration. The three phases of cough sound signals are shown in Fig.1. The Phase-1 of cough signal is the result of airway narrowing and bifurcations, leading to turbulent airflow resulting in a higher amplitude and shorter duration as compared to the other two phases. Phase-2 is the result of airflow in trachea and collection of mucus in the trachea, which results in longer duration and lower amplitude than Phase-1 and Phase-3. Phase-3 is vocal fold adduction, which appears as decaying signal as airflow reduces. This Phase-3 may not appear in some cases [8]. The cough sound signals will have frequency distribution from 350Hz to 4KHz [9].

*A. MFCC (Mel Frequency Cepstral Coefficents)*

The human auditory perception is modeled in the form of logarithmic curve, where log of power spectrum, which is cepstrum gives perceptual sensitivity of a human along magnitude axis and the other frequency axis is perceptual scale judged by humans in the form of Mel-scale. The mapping of frequency to Mel-scale is approximately linear below 1KHz and logarithmic for higher frequencies. The Mel scale conversion and vice versa are as below.

$$m = 2595 \, log_{10}(1 + f/700)$$
$$f = 700(10^{m/2595} - 1) \quad (1)$$

Applying triangular filter banks, helps in weighted sum of energies near the target frequency which helps in robust estimate due to energy integration along with applying perceptual frequency scale. This form of weighted and perceptual frequency scaled coefficients are called Mel frequency cepstral coefficient (MFCC) as below :

$$mfcc_i = \sqrt{2/N} \sum_{j=1}^{N} log(x_j) \cos(i\pi(j - 0.5)/N) \quad (2)$$

Where $mfcc_i$ is i-th Mel Frequency Cepstral Coefficient and $x_j$ is the j-th Mel-Filter bank's energy.

MFCC is chosen as the feature since it quantifies the gross shape of the spectrum (the spectral envelope), which is important in detections of macro level information of Phase-1 or Phase-2 of cough events. It is computationally efficient and well tested and understood. The Mel filters applied can be chosen smartly for signals other than cough events, which can be discarded by choosing the appropriate coefficients. With these advantages, the MFCC as an input feature for neural network is chosen with different parameters, as described in part-IV.

MFCC applied on waveform from Fig.1, with hop length of 5 msec, is as shown in Fig.2. The colored intensities show variation of frequency and their log spectrum very clearly. For this interpretation, the MFCC which is of (number of mfcc, number of frames) dimension must be converted into 3channel RGB color space of dimension (number of mfcc, number of frames,3channels), which may add memory space and computational cost.

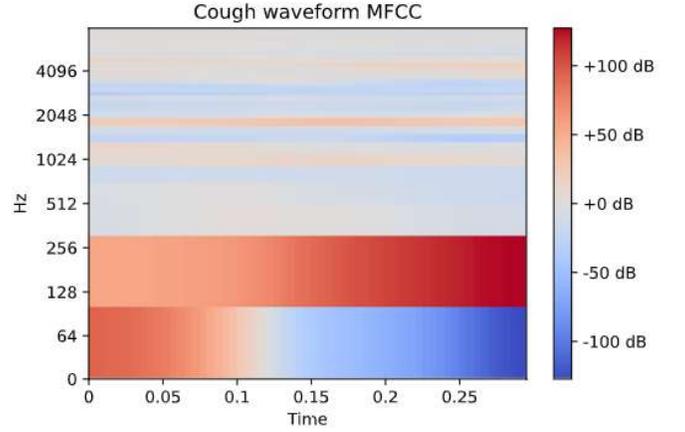

Fig. 2: Cough waveform MFCC, with colored intensity variations showing frequency distribution and amplitude levels.

The MFCC from color space would be a better representation for neural network to learn just like computer vision images. To implement the feature in constrained devices, the paper retains the MFCC as it is. Hence, what the neural network sees is as below in Fig.3.

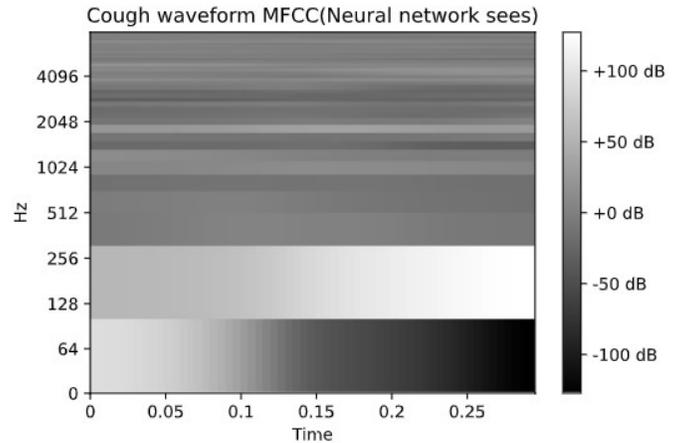

Fig. 3: Cough waveform MFCC, what Neural network sees.

III. ALGORITHM DEVELOPMENT FOR COUGH DETECTION

The event of cough detection is so sensitive as its frequency of distribution falls very close to speech frequency distribution. Hence to make it more robust in event detection, the Convolutional Neural network(CNN) approach is chosen for the core part of detection, as shown in Fig.4.

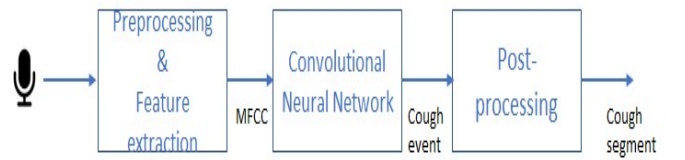

Fig. 4: System architecture

The microphone sound signal is sampled at 16KHz, with 16-bit PCM as input to the system. The preprocessing consists of Gain control, Band pass filter and Onset detections. The Onset events will help to know where the onset spells of cough or significant audio signal energy has occurred. The feature extractor is MFCC which will transform the audio signal into Mel-cepstral domain of the form shown in Fig.3. This image with dimension of (Number of MFCC, Number of frames), will feed into light Convolutional neural network, which can detect occurrence of cough event. The audio signal is segmented in to 1sec duration worth of data. Hence the cough events occurring at the boundaries and carrying to next segments are properly consolidated using post processing.

### A. Pre-processing and Feature extraction

The Automatic gain control and noise removal in the form of Band pass filter are added as preprocessing step. The paper depends on Phase-1 characteristics of the cough sound signal, whose frequency of significant energy is in the range of 150 to 2KHz as shown in Fig.5. Thus, this frequency band is selected as pass band for band pass filter.

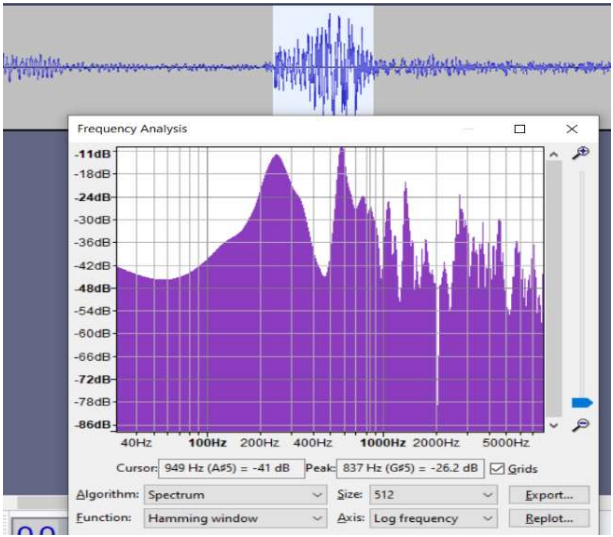

Fig. 5: Phase-1 waveform and corresponding frequency spectrum using Audacity tool.

The onset detection, which is part of pre-processing will ensure that the model is not triggered when there is silence or noise. The onset works and detects based on sudden burst of energy, a change in the short-time spectrum of the signal or in the statistical properties, etc. The onset strength is calculated based on superflux detection function [10]. This strength is then peak picked based on strength greater than 0.5. Since super flux also uses logarithmic based spectrum, there is saving of computational cost and memory, as MFCC is calculated on the same basis.

The MFCC description as given in Part-II, is used as feature extractor here. With onset peak picking, the signal of audio is fed in to MFCC module, which derives 40 coefficients, with Number of frames divided over 1 second duration of segment. The analysis or details of number of frames which basically segments into short time duration are described in Part-IV. As the audio segment is expected to be 1sec duration, any short fall of this, either due to onset spell or smaller duration audio less than 1sec, is compensated by padding zeros into the audio, thus ensuring MFCC of fixed dimension.

### B. Convolutional Neural network

To target a low-cost Microcontroller class based smart devices, the neural network is chosen carefully based on convolution neural network. The Fig.6 below shows the summary (using keras framework) of the architecture of convolutional neural network for cough detection.

```
Model: "model"
_________________________________________________________________
Layer (type)                 Output Shape              Param #
=================================================================
input_1 (InputLayer)         [(None, 40, 267, 1)]      0
_________________________________________________________________
conv2d (Conv2D)              (None, 19, 133, 16)       160
_________________________________________________________________
dropout (Dropout)            (None, 19, 133, 16)       0
_________________________________________________________________
conv2d_1 (Conv2D)            (None, 9, 66, 32)         4640
_________________________________________________________________
dropout_1 (Dropout)          (None, 9, 66, 32)         0
_________________________________________________________________
conv2d_2 (Conv2D)            (None, 4, 32, 40)         11560
_________________________________________________________________
dropout_2 (Dropout)          (None, 4, 32, 40)         0
_________________________________________________________________
global_max_pooling2d (Global (None, 40)                0
_________________________________________________________________
batch_normalization (BatchNo (None, 40)                160
_________________________________________________________________
dense (Dense)                (None, 1)                 41
=================================================================
Total params: 16,561
Trainable params: 16,481
Non-trainable params: 80
```

Fig. 6: Convolution neural network for cough detection.

The input dimension shown here is MFCC of 40 coefficients and 5msec frames with 25% overlap (making hop length of 5*0.25=3.75msec), which were divided among 1sec segment constituting ceil(1000/3.75) = 267 frames. The 3-2D-convolutional layers are used, with max pooling to pick the best feature after convolution operations. The Batch normalization normalizes the features and then Dense layer of small dimension is used to reduce the load (This operation requires matrix multiplication, which can be heavier) on CPU cycles, followed up with sigmoid operation to classify the detection. As seen from Fig.6, the number of parameters is 16561, which can take 65KB of weights memory with float implementation. The Quantized model will lead us to 16.25KB with int8 weights.

### C. Post processing

The cough audio signal is segmented in to 1 sec interval for inferring detection of cough events. This brings possibility of cough events occurring at the boundary, with continuation into immediate next segment, showing them as two different segments. To mitigate such problems, the post processing will take the input from onset detection and inference output to make judgment of continuity in cough events.

As an example, Fig.7 shows the two segments of the cough, where the continuity of cough is spilled over to the next segment. The 1st segment is from 0.3sec to 1.3 sec. The 2nd segment onset spell occurs at 1.35sec, which is cross checked with 1st segment Phase-3 or Phase-1 onset along with the inference. After the 2nd segment inference, the result is consolidated by knowing the onset strength points, to end at 1.8 sec.

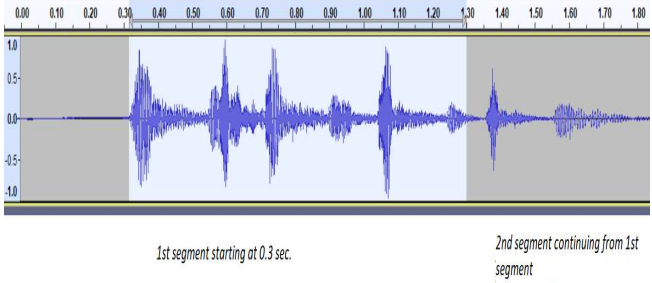

Fig. 7: Cough segments going beyond 1 sec intervals.

## IV. PERFORMANCES AND RESULTS

This section will give details about parameters affecting the results and eventually leading to performance of the neural network.

### A. Database and their usage

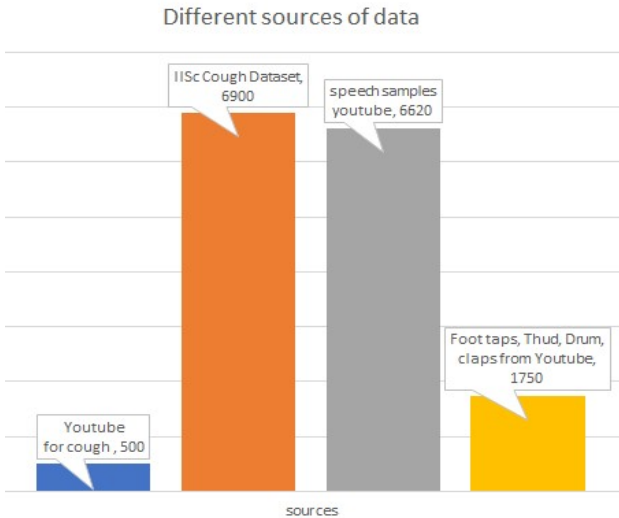

Fig. 8: Different sources of data used in training CNN.

The CNN network is trained with dataset retrieved from sources on YouTube and IISc Coswara datasets [11]. Out of 7400 cough samples (each of 1sec duration), 6900 in number are used from IISc datasets, as shown in Fig.8. The YouTube source is used to get few cough samples and speech samples along with foot tap, thud and drum, claps sound. The YouTube samples are segmented in to 1 sec and manually labelled for cough segments and resampled to 16KHz.

The paper used 72%, 8% and 20% as a training, validation, and testing data, respectively. The results reported here are from testing samples. The stratified sampling mode is used, so that there is uniform distribution of samples from both cough and unknown classes.

### B. MFCC parameters, relating to cough characteristics

One of the MFCC parameter which plays important role are the frame length and hop length. These lengths can be chosen as per the cough characteristic, which can help neural network to learn faster and with better accuracy.

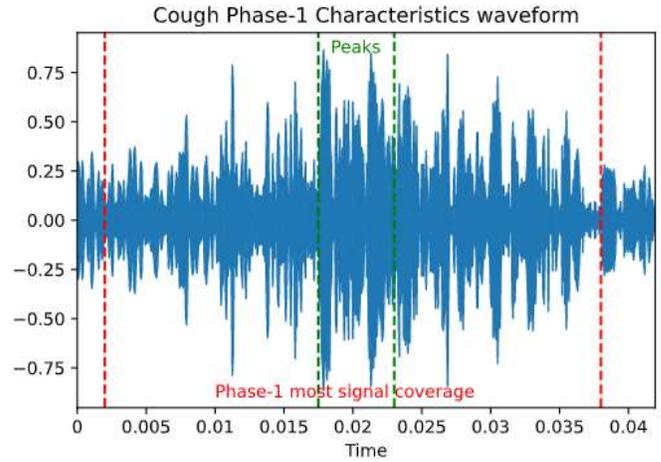

Fig. 9: 5msec and 35 msec segments of cough from phase 1.

| Frame length (ms) | overlap(%) | Hop length (ms) | No.of frames | Total MFCC coefficents (40x No.of frames) | Memory for MFCC (KB) | Pre-processing Memory (KB) |
|---|---|---|---|---|---|---|
| 5 | 0 | 5 | 200 | 8000 | 31.25 | 39.65 |
| 5 | 25 | 3.75 | 267 | 10680 | 41.71875 | 50.11875 |
| 20 | 0 | 20 | 50 | 2000 | 7.8125 | 16.2125 |
| 20 | 25 | 15 | 67 | 2680 | 10.46875 | 18.86875 |
| 35 | 0 | 35 | 29 | 1160 | 4.53125 | 12.93125 |
| 35 | 25 | 26.25 | 39 | 1560 | 6.09375 | 14.49375 |
| 50 | 0 | 50 | 20 | 800 | 3.125 | 11.525 |
| 50 | 25 | 37.5 | 27 | 1080 | 4.21875 | 12.61875 |
| 70 | 0 | 70 | 15 | 600 | 2.34375 | 10.74375 |
| 70 | 25 | 52.5 | 20 | 800 | 3.125 | 11.525 |

Table-1: Memory and MFCC parameter variations

As shown in Fig.9, the peaks occur in those 5msec of segments, whereas 70% of phase-1 gets covered with 35msec waveform. The paper demonstrates that these two-frame lengths as the best parameters, which helps in cough detection with highest F1-score. The Table 1 shows the memory picture for onset strength and MFCC in preprocessing memory column, where a scratch memory of 8.4KB is considered for processing. The 5msec with 25% overlap shows highest F1-score (98.17%), with higher memory of 50KB. Whereas 35ms with 0% overlap, shows good balance between F1-Score(97.27%) and Memory of 12.8KB. The paper shows various other frame length performances in Fig.10. The 70 ms which covers Phase-1 of cough characteristics, have better F1-score but their negative predictive value (NPP) and sensitivity(SE) are poor.

## C. Neural network parameters

The Neural network hyper parameters used have learning rate (lr) of **0.005**, optimizer used is **Nadam**, with number of epochs kept up to 200. The learning rate plateau, where the learning rate is reduced by factor of 10 is applied to fine tune the convergence if the loss does not improve beyond 10 epochs. The neural network architecture shown in Fig.6 learns and saturates around 65-80 epochs, depending on feature of different parameter. The best validation loss is picked for inference model.

## D. Performance metric

The performance of cough detection is measured in terms of sensitivity, specificity, positive predictive value (PPV), negative predictive value (NPV) and F1-score. The definitions of these metrics are:

$$SE = \frac{TP}{TP + FN} \qquad SP = \frac{TN}{TN + FP}$$

$$PPV = \frac{TP}{TP + FP} \qquad NPV = \frac{TN}{TN + FN}$$

$$F1 = 2 * \frac{SE * PPV}{SE + PPV} \qquad (3)$$

TP is True Positive  : correctly detected cough event samples.
TN is True Negative  : correctly detected non-cough samples.
FP is False Positive  : incorrectly labelled non-cough samples.
FN is False Negative :  incorrectly missed cough event samples.

The Sensitivity and Specificity are the measures of true positive rate and true negative rates. The PPV and NPV are the measures of prevalence. PPV is also called as Precision. The NPV is probability of how best the non-cough or unknown events are detected. F1score covers sensitivity and PPV, giving measure of balance between the two. In the results section the paper emphasizes on Sensitivity, Specificity and NPV with F1-score, as PPV is covered inside F1-score.

## E. Results

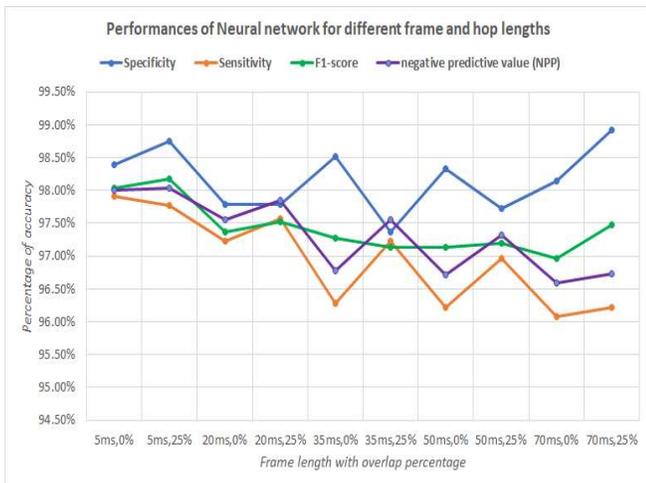

Fig. 10: Neural network performance for different frame lengths.

The Fig.10 shows the graph of performance metric for different parameters on testing dataset which is of 20% of data shown in Fig.8.

The CNN network for cough detection is estimated to be around 13.26Million Macs(Multiply and Accumulate), 3.26K Additions, which is suitable for microcontroller class of 50MHz and 200KB RAM memory, enabling patient monitoring system of lesser cost.

## V. CONCLUSIONS

In this paper, the cough detection classification network is designed and developed using CNN, which is of 16K parameters, and can easily fit into tiny IoT based smart devices for patient monitoring systems. The paper shows selection of MFCC feature, with its parameters playing critical role in getting best performance at reasonable memory space. The paper shows F1-score of 98.17% with memory up to 50KB for preprocessing, 65KB for Neural network weights.

## REFERENCES

[1] R. X. Adhi Pramono, S. Anas Imtiaz and E. Rodriguez-Villegas, "Automatic Identification of Cough Events from Acoustic Signals," 2019 41st Annual International Conference of the IEEE Engineering in Medicine and Biology Society (EMBC), Berlin, Germany, 2019, pp. 217-220, doi: 10.1109/EMBC.2019.8856420.

[2] I. S. Jacobs and C. P. Bean, "Fine particles, thin films and exchange anisotropy," in Magnetism, vol. III, G. T. Rado and H. Suhl, Eds. New York: Academic, 1963, pp. 271–350.

[3] J. Amoh and K. Odame, "Deep neural networks for identifying cough sounds," IEEE transactions on biomedical circuits and systems, vol. 10, no. 5, pp. 1003–1011, 2016.

[4] Liu, Jia-Ming & You, Mingyu & Wang, Zheng & Li, Guo-Zheng & Xu, Xianghuai & Qiu, Zhongmin. (2014). Cough detection using deep neural networks. Proceedings - 2014 IEEE International Conference on Bioinformatics and Biomedicine, IEEE BIBM 2014. 560-563. 10.1109/BIBM.2014.6999220.

[5] Y. Yorozu, M. Hirano, K. Oka, and Y. Tagawa, "Electron spectroscopy studies on magneto-optical media and plastic substrate interface," IEEE Transl. J. Magn. Japan, vol. 2, pp. 740–741, August 1987 [Digests 9th Annual Conf. Magnetics Japan, p. 301, 1982].

[6] Y. A. Amrulloh, U. R. Abeyratne, V. Swarnkar, R. Triasih, and A. Setyati, "Automatic cough segmentation from non-contact sound recordings in pediatric wards," Biomedical Signal Processing and Control, vol. 21, pp. 126–136, 2015.

[7] G.A. Fontana, "Before we get started: what is a cough?," lung, supp. 1, s3-6, Oct. 2008. (DOI 10.1007/s00408-007-9036-8)

[8] H. Chatrzarrin, A. Arcelus, R. Goubran and F. Knoefel, "Feature extraction for the differentiation of dry and wet cough sounds," 2011 IEEE International Symposium on Medical Measurements and Applications, Bari, Italy, 2011, pp. 162-166, doi: 10.1109/MeMeA.2011.5966670.

[9] S.H. Shin, T. Hashimoto, and S. Hatano, "Automatic detection system for cough sounds as a symptom of abnormal health condition," IEEE Transaction on Information Technology in Biomedicine, vol. 13, no. 4, Jul. 2009.

[10] Böck, Sebastian, and Gerhard Widmer. "Maximum filter vibrato suppression for onset detection." 16th International Conference on Digital Audio Effects, Maynooth, Ireland. 2013.

[11] https://github.com/iiscleap/Coswara-Data